\documentclass[10pt, conference, compsocconf]{IEEEtran}

\usepackage{graphicx}
\usepackage{hhline,theorem,verbatim,euscript,url}
\usepackage[babel=true]{csquotes}
\usepackage{listings}
\usepackage{amssymb}
\usepackage{amstext}
\usepackage{mathtools}
\usepackage{textcomp}
\usepackage{array}
\usepackage{float} 
\usepackage{subfigure}
\usepackage{color}
\usepackage{textcomp}
\usepackage{fancyvrb}
\usepackage[abs]{overpic}
\definecolor{dkgreen}{rgb}{0,0.6,0}
\definecolor{gray}{rgb}{0.5,0.5,0.5}
\definecolor{mauve}{rgb}{0.58,0,0.82}

\newtheorem{theorem}{Theorem}
\newtheorem{definition}[theorem]{Definition}
\newtheorem{proposition}[theorem]{Proposition}

\DeclareMathOperator{\Verdict}{Verdict} 
\DeclareMathOperator{\defense}{defense} 
\DeclareMathOperator{\Vulnerable}{Vulnerable}
\DeclareMathOperator{\Unsat}{Unsat}
\DeclareMathOperator{\Inconclusive}{Inconclusive}

\DeclareMathOperator{\BADStep}{BADStep}

\begin{document}
\title{An Advanced Approach for Choosing Security Patterns and Checking their Implementation}

\author{\IEEEauthorblockN{S\'ebastien Salva\IEEEauthorrefmark{1}, Loukmen Regainia\IEEEauthorrefmark{2}
	\IEEEauthorblockA{LIMOS - UMR CNRS 6158\\
	University Clermont Auvergne, France\\		
	Email: \IEEEauthorrefmark{1} sebastien.salva@uca.fr,
	\IEEEauthorrefmark{2} loukmen.regainia@uca.fr}}}

% make the title area
\maketitle

\begin{abstract}

This paper tackles the problems of generating concrete test cases for testing whether an application is vulnerable to attacks, and of checking whether security solutions are correctly implemented. The approach proposed in the paper aims at guiding developers towards the implementation of secure applications, from the threat modelling stage up to the testing one. This approach relies on a knowledge base integrating varied security data, e.g., attacks, attack steps, and  security patterns that are generic and re-usable solutions to design secure applications. The first stage of the approach consists in assisting developers in the design of Attack Defense Trees expressing the attacker possibilities to compromise an application and the defenses that may be implemented. These defenses are given under the form of security pattern combinations. In the second stage, these trees are used to guide developers in the test case generation. After the test case execution, test verdicts show whether an application is vulnerable to the threats modelled by an ADTree. The last stage of the approach checks whether behavioural properties of security patterns hold in the application traces collected while the test case execution. These properties are formalised with LTL properties, which are generated from the knowledge base. Developers do not have to write LTL properties not to be expert in formal models. We experimented the approach on 10 Web applications to evaluate its testing effectiveness and its performance.

\end{abstract}

\section{Introduction}
\label{sec:intro}

Today’s developers are no longer just expected to code and build applications. They also have to ensure that applications meet minimum reliability guarantees and security requirements. Unfortunately, choosing security solutions or testing software security are not known to be simple or effortless activities. Developers are indeed overloaded of new trends, frameworks, security issues, documents, etc. Furthermore, they sometimes lack skills and experience for choosing security solutions or writing concrete test cases. They need to be guided on how to design or implement secure applications and test them, in order to contribute in a solid quality assurance process.

This work focuses on this need and proposes an approach that guides developers devise more secure applications from the threat modelling stage, which is a process consisting in identifying the potential threats of an application, up to the testing one. The present paper is an extended version of \cite{RS18}, which provides additional  details  on the security test case generation, the formalisation of behavioural properties of security patterns with Linear Temporal Logic (LTL) properties, and on their automatic generation. We also provide an evaluation of the approach and discuss the threats to validity. 

%This work brings together the notions of security documents, threat modelling, the writing of concrete test cases and their executions to help developers in these tasks. 

In order to guide developers, our approach is based upon several several digitalised security bases or documents gathered in a knowledge base. In particular, the latter includes security solutions under the form of security patterns, which can be chosen and applied as soon as the application design. Security patterns are defined as \textit{reusable elements  to design secure applications, which will enable software architects and designers to produce a system that meets their security requirements and that is maintainable and extensible from the smallest to the largest systems} \cite{Rodriguez2003}. Our approach helps developers chose security patterns with regard to given security threats. Then, it builds security test cases to check whether an application is vulnerable, and test whether security patterns are correctly  implemented in the application. More precisely, the contributions of this work are summarised in the following points:

\begin{itemize}
	%	\item Data acquisition and integration: the first part of our method and tool aims to semi-automatically extract knowledge from various Web and publicly accessible sources to conceive a data-store storing relationships among attacks, attack steps, techniques, security principles, security patterns and test case parts, which we call sections, written with the Given When Then (GWT) template. We have chosen to express security solutions with security patterns as these are well documented in the literature and in catalogues. These solutions, described with texts or UML diagrams, aim at implementing into the application counter-measures to threats or attacks.	As for the GWT test case formulation, it offers better readability and re-usability by separating a test case into several sections themselves associated to procedures implementing them. 
	%Our method associates attack steps and security patterns with Given, When, Then test case sections and procedures;
	
	\item the approach assists developers in the threat modelling stage by helping in the generation of Attack Defense Trees (ADTrees) \cite{kordy2012attack}. The latter express the attacker possibilities to compromise an application, and give the defenses that may be put in place to prevent attacks. Defenses are here expressed with security patterns. We have chosen this tree model because it offers the advantage of being easy to understand even for novices in security; 
	
	%\item Security pattern choice: the developer can edit the previous ADTrees to add or remove undesired attack or steps. He or she also has to choose security patterns so that it remains pattern conjunctions only in the ADTrees. Our generated ADTrees are conceived to guide him or her in this task. Indeed, patterns are combined with classical logic operations (and, or). In addition, they provide inter-pattern relation-ships. For instance, ADTrees show the conflicting or dependent patterns. Our method finally provides a final ADTree derived from the composition of the initial one with these ADtrees. This last model may be used as security requirement documents for threat risk modelling;
	
	\item the second part of the approach supports developers in writing concrete security test cases. A test suite is automatically extracted from an ADTree. The test suite is made up of test case stubs, which are completed with comments or blocs of code. Once completed, these are used to experiment an application under test (shortened $AUT$), seen as a black-box. The test case execution provides verdicts expressing whether the $AUT$ is vulnerable to the threats modelled in the ADTree;

	%When an application is not vulnerable to an attack step (expressed by a test verdict) then we deduce, , ;
	
	\item the last part of the approach allows developers to check whether security patterns are correctly implemented in the application. Kobashi et al. dealt with this task by asking users to manually translate security pattern behaviours into formal properties \cite{Kobashi15}. Unfortunately, few developers have the required skills in formal modelling. We hence prefer proposing a practical way to generate them.  After the security pattern choice, our approach provides generic UML sequence diagrams, which can be adapted to better match the application context. From these diagrams, the approach automatically  generate  LTL properties. After the test case execution, we check if these properties hold in the application traces. The developer is hence not aware of the LTL property generation.  
	
\end{itemize}

We have implemented this approach in a tool prototype available in \cite{data}. This tool was used to conduct several experiments on 10 Web applications to evaluate the security testing and security pattern testing effectiveness of the tool as well as its performance.

%Several digitalised security bases, documents and papers have been proposed to guide developers in these activities. For instance, the Common Attack Pattern Enumeration and Classification (CAPEC) makes publicly available around 1000 attack descriptions, including their goals, steps, techniques, the targeted vulnerabilities, etc. 

\subsubsection*{Paper Organisation}  Section \ref{sec:background} outlines the context of this work. We recall some basic concepts and notations about security patterns and ADTrees. We also discuss about  the related work and our motivations. Section \ref{sec:datastore} briefly presents the architecture of the knowledge base used by our approach.  The approach steps are described in Section \ref{sec:app}. These steps are gathered into 3 stages called threat modelling, security testing, and security pattern testing. Subsequently, Section \ref{sec:impl} describes our prototype implementation, and Section \ref{sec:eval} evaluates the approach. Finally,  Section \ref{sec:conclusion} summarizes our contributions and presents future work.

\section{Background}
\label{sec:background}

This section recalls the basic concepts related to security patterns and Attack Defense trees. The related work is presented thereafter.

\subsection{Security Patterns}

Security patterns provide guidelines for secure system design and evaluation \cite{Yoder1998}. They also are considered as countermeasures to threats and attacks \cite{Schumacher2003}. Security patterns have to be selected in the design stage, integrated in application models, and eventually implemented. Their descriptions are usually given with texts or schema. But, they are often characterised by UML diagrams capturing structural or behavioural properties.

%In another context, security pattern catalogues, e.g., \cite{patrepo}, list 176 re-usable solutions for helping developers design more secure applications. The security pattern, which is a topic of this paper, \textit{intuitively relates countermeasures to threats and attacks in a given context} \cite{Schumacher2003}. This profusion of documents makes developers drown in a sea of recommendations taking security with different viewpoints (attackers, defenders, etc.), abstraction levels (security principles, attack steps, exploits, etc.) or contexts (system, network, etc.). %As developers cannot be experts in any software engineering fields, these difficulties entail that software security is not always well performed. 

Several security pattern catalogues are available in the literature, e.g., \cite{patrepo,Yskout2015}, themselves extracted from other papers. In these catalogues, security patterns are systematically organised according to features and relationships among them. Among these features, we often find the solutions called intents, or the interests called forces. A security pattern may have different relationships with other patterns. These relations may noticeably help combine patterns together and not to devise unsound composite patterns. Yskout et al.  proposed the following annotations between two patterns \cite{yskout2006system}: \enquote{depend}, \enquote{benefit}, \enquote{impair} (the functioning of the pattern can be obstructed by the implementation of a second one), \enquote{alternative}, \enquote{conflict}.

\begin{comment}
\item $p_1$ depend $p_2$  means that the implementation of $p_1$ requires the implementation of $p_2$; 

\item $p_1$ benefit $p_2$ expresses that implementing $p_2$ completes $p_1$ with extra security functionalities or decreases the development time. However, $p_1$ can be correctly implemented despite the absence of $p_2$;

\item $p_1$ impair $p_2$ means that the functioning of $p_1$ can be obstructed by the implementation of $p_2$;%, but both may be used together with care;

\item $p_1$ alternative $p_2$ expresses that $p_2$ is a different pattern fulfilling the same functionality as $p_1$;

\item $p_1$ conflict $p_2$ encodes the fact that if both $p_1$ and $p_2$ are implemented together then it shall result in inconsistencies.
\end{comment}

\begin{figure}
	\centering
	\includegraphics[width=1\linewidth]{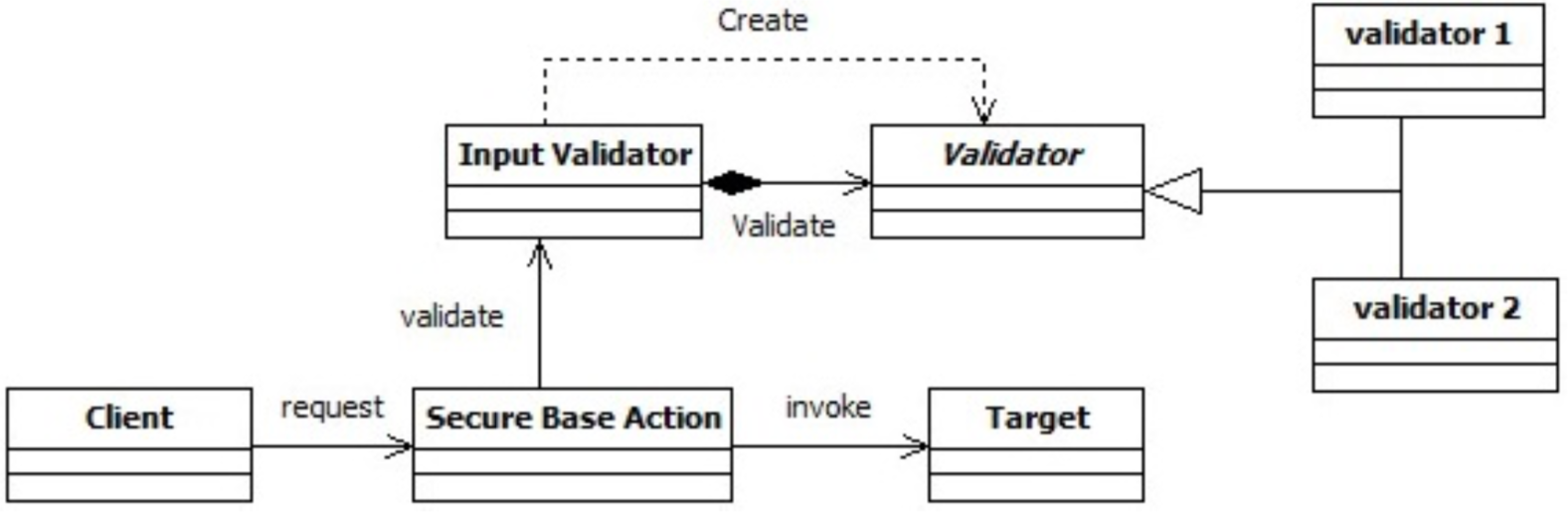}
	\caption[]{Class layout of the security pattern \enquote{Intercepting Validator}.}
	\label{fig:1}
\end{figure}

Figure \ref{fig:1} %portrays 
depicts the UML structural diagram of the security pattern \enquote{Intercepting Validator}, which is taken as example is the remainder of the paper.  Its purpose is to provide the application with a centralized validation mechanism, which applies some filters (Validator classes) declaratively based on URL, allowing different requests to be mapped to different filter chains. This validation mechanism is decoupled from the other parts of the application and each data supplied by the client is validated before being used. The validation of input data prevents attackers from passing malformed input in order to inject malicious commands.
 
\subsection{Attack Defense Trees}

ADTrees \textit{are graphical representations of possible measures an attacker might take in order to compromise a system and the defenses that a defender may employ to protect the system} \cite{kordy2012attack}. ADTrees have two different kinds of nodes: attack nodes (red circles) and defense nodes (green squares). A node can be  \textit{refined} with child nodes and can have one child of the opposite type (linked with a dashed line). Node refinements can be disjunctive  or conjunctive. The former is recognisable by edges going from a node to its children. The latter is graphically distinguishable by connecting these edges  with an arc. We extend these two refinements with the sequential conjunctive refinement of attack nodes, defined by the same authors in \cite{jhawar2015attack}. This operator expresses the execution order of child attack nodes. Graphically, a sequential conjunctive refinement is depicted by connecting the edges, going from a node to its children, with an arrow. 

\begin{figure}
	\centering
	\includegraphics[width=.5\linewidth]{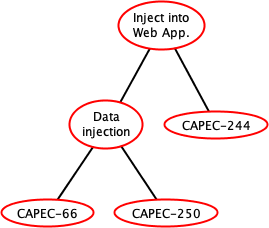}
	\caption[]{ADTree example modelling injection attacks}
	\label{fig:injection}
	%\vspace{-0.2cm}
\end{figure}

For instance, the ADTree of Figure \ref{fig:injection} identifies the objectives of an attacker or the possible vulnerabilities related to the supply of untrusted inputs to an application. The root node is here detailed with disjunctive refinements connecting three leaves, which are labelled by attack referenced in a base called the Common Attack Pattern Enumeration and Classification (CAPEC) \cite{CAPEC}. The node CAPEC-66 refers to \enquote{SQL Injection}, CAPEC-250 refers to XML injections and CAPEC-244 to \enquote{Cross-Site Scripting via Encoded URI Schemes}.

An ADTree $T$ can be formulated with an algebraic expression called ADTerm and denoted $\iota(T)$. In short, the ADTerm syntax is composed of operators having types given as exponents in $\{o,p\}$ with $o$ modelling an opponent and $p$ a proponent. $\vee^s, \wedge^s, \overrightarrow{\wedge}^s$, with $s \in \{o,p\}$ respectively stand for the disjunctive refinement, the conjunctive refinement and the sequential conjunctive refinement of a node. A last operator $c$ expresses counteractions (dashed lines in the graphical tree). $c^s(a,d)$ intuitively means that there exists an action d (not of type s) that counteracts the action a (of type s).
The ADTree of Figure \ref{fig:injection} can be represented with the ADTerm $\vee^p ( \vee^p ($CAPEC-66, CAPEC-250$),$ CAPEC-244$)$.

\subsection{Related Work}
The literature proposes several papers dealing with the test case generation from  Attack trees (or related models) and some other ones about security pattern testing. As these topics are related to our work, we introduce them below and give some observations.

\subsubsection{Security Testing From Threat Models}
the generation of concrete test cases from models has been widely studied in the last decade, in particular to test the security level of different kinds of systems, protocols  or software. Most of the proposed approaches take specifications expressing the expected behaviours of the implementation. But, other authors preferred to bring security aspects out and used models describing attacker goals or vulnerability causes of the system. Such models are conceived during the threat modelling phase of the system \cite{threatmodeling}, which is considered as a critical phase of the software life cycle since \textit{"you cannot build a secure system until you understand your threats!"} \cite{securecode}. Schieferdecker et al.  presented a survey paper referencing some approaches in this area \cite{MdlBST}. For instance, Xu et al. proposed to test the security of Web applications with models as Petri nets to describe attacks \cite{Xu12}. Attack scenarios are extracted from the Reachability graphs of the Petri nets. Then, test cases written for the Selenium tool are generated by means of a MIM (Model- Implementation Mapping) description, which maps each Petri net place and transition to a block of code. Bozic et al. proposed a security testing approach associating UML state diagrams to represent attacks, and combinatorial testing to generate input values used to make executable test cases derived from UML models \cite{Bozic2014}.

Other authors adopted models as  trees (Attack trees, vulnerability Cause Graphs, Security Activity Graphs, etc.) to represent the threats, attacks or vulnerability causes that should be prevented in an application. From these models, test cases are then written to check whether  attacks can be successfully executed or whether vulnerabilities are detected in the implementation. Morai et al. introduced a security testing approach specialised for network protocols \cite{Morais2009}. Attack scenarios are extracted from an Attack tree and are converted to Attack patterns and UML specifications. From these, attack scripts are manually written and are completed with the injection of (network) faults. %The security testing method proposed in  \cite{Marback09} shares some similarities with \cite{Xu12}: 
In the security testing method proposed in \cite{Marback09}, data flow diagrams are converted into Attack trees from which sequences are extracted. These sequences are composed of events combined with parameters related to regular expressions. These events are then replaced with blocks of code to produce test cases. The work published in \cite{ElAriss2011} provides a manual process composed of eight steps. Given an Attack tree, these steps transform it into a State chart model, which is iteratively completed and transformed before using a model-based testing technique to generate test cases. In \cite{Marback2013}, test cases are generated from Threat trees. The latter are previously completed with parameters associated to regular expressions to generate input values. Security scenarios are extracted from the Threat trees and are manually converted to executable test scripts. Shahmehri et al. proposed a passive testing approach, which monitors an $AUT$ to detect vulnerabilities \cite{Shahmehri2012}. The undesired vulnerabilities are modelled with security goal models, which are specialised directed acyclic graphs showing security goals, vulnerabilities and eventually mitigations. Detection conditions are then semi-automatically extracted and given to a monitoring tool.  

We observed that the above methods either automatically generate abstract test cases from (formal) specifications or help write concrete test cases from  detailed threat models. On the one hand, as abstract test cases cannot be directly used to experiment an $AUT$, some works proposed test case transformation techniques. However, this kind of technique is at the moment very limited. On the other hand, Only a few of developers have the required skills to write threat models or  test cases, as  a strong expertise on security is often required. Besides, the methods neither guide developers in the threat modelling phase nor provide any security solution. We focused on this problem and laid the first stone of the present approach in \cite{SR17a,SR17b,salva:hal-02019145}. We firstly presented a semi-automatic data integration method \cite{SR17a} to build security pattern classifications.  This method extracts security data from various Web and publicly accessible sources and stores relationships among attacks, security principles and security patterns into a knowledge base. Section \ref{sec:datastore} summarises the results of this work used in this paper, i.e., the first meta-model version of the  data-store. In \cite{SR17b}, we proposed an approach to help developers write ADTrees and concrete security test cases to check whether an application is vulnerable to these attacks. This work was extended in \cite{salva:hal-02019145} to support the generation of test suites composed of lists of ordered GWT test cases, a list being devoted to check whether an AUT is vulnerable to an attack, which is segmented into an ordered sequence of attack steps. This test suite organisation is used to reduce the test costs with the deduction of some test verdicts under certain conditions. However, it does not assist developers to ensure that security patterns have been correctly implemented in the application. This work supplements our early study by covering this part.

\subsubsection{Security Pattern Testing}

the verification of patterns on models was studied in \cite{Dong2010,Hamid2012,Yoshizawa14,Kobashi15,RBS16}. In these papers, pattern goals or intents or structural properties are specified with UML sequence diagrams \cite{Dong2010} with expressions written with the Object Constraint Language (OCL) \cite{Hamid2012,Yoshizawa14,Kobashi15} or with LTL properties \cite{RBS16}. The pattern features are then checked on UML models. %These works target the verification of As security pattern properties are not tested on applications under test in these papers, we consider outside the scope of this paper.

% For instance, Dongs et al. proposed to verify the compositions of security patterns using model checking \cite{Dong2010}. They introduced guidelines to specify security pattern behaviours with UML sequence diagrams, which are transformed into specifications written with the CSS process algebra. Then, a model-checker is called to check whether these specifications are met on the model of the system. Like in the previous approach, we proposed in \cite{RBS16} to guide designers for checking whether a set of security patterns is correctly integrated into UML models. But, we considered the structural properties of security patterns. Besides, our approach also checks whether a UML model exposes vulnerabilities despite the use of security patterns. In these works, the implementation is not considered nor tested. 

%out of the scope ???

Few works dealt with the testing of security patterns, which is the main topic of this paper. Yoshizawa et al. introduced a method for testing whether behavioural and structural properties of patterns may be observed in application traces \cite{Yoshizawa14}. Given a security pattern, two test templates (Object Constraint Language (OCL) expressions) are manually written, one to specify the pattern structure and another one to encode its behaviour. Then, developers have to make templates concrete by manually writing tests for experimenting the application. The latter returns traces on which the OCL expressions are verified. 

We observed that these previous works require the modelling of security patterns or vulnerabilities with formal properties. Instead of assuming that developers are expert in the writing of formal properties, we propose a  practical way to generate them. Intuitively, after the choice of security patterns, our approach provides generic UML sequence diagrams, which can be modified by a developer. From these diagrams, we automatically generate LTL properties, which capture the cause-effects relations among pairs of method calls. After the test case execution, we check if these properties hold in the  application traces, obtained while the test case execution. The developer is hence not aware of the LTL property generation. As stated in the introduction, this work provides more details on test case generation and on the formalisation of behavioural properties of security patterns with LTL properties. We also complete the transformation rules allowing to derive more LTL properties from UML sequence diagrams. We also provide an evaluation of the approach targeting the security pattern testing stage and discuss the threats to validity.  

\section{Knowledge Base Overview}
\label{sec:datastore}

Our approach relies on a knowledge base, denoted KB in the remainder of the paper. It gathers information allowing to help or automate some steps of the testing process. We summarise its architecture in this section but we refer to \cite{SR17a} for a complete description of its associations and of the data integration process.

\subsection{Knowledge Base Meta-Model}

\begin{figure*}[ht]
	\centering
	\includegraphics[width=.7\linewidth]{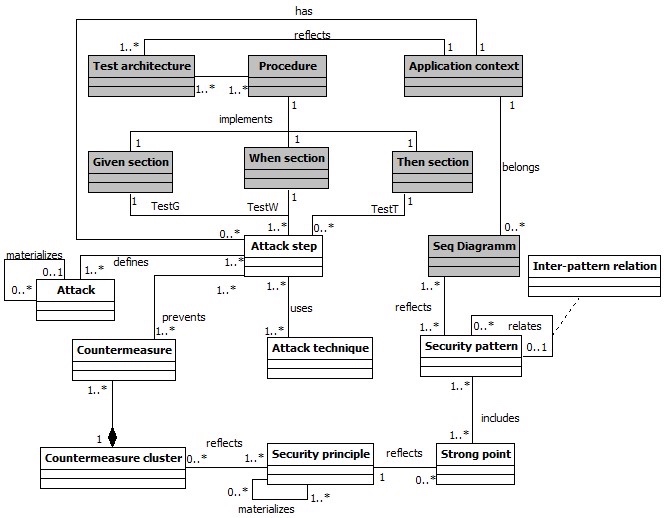}
	\caption{Data-store meta-model}
	\label{fig:datastore1}
	%\vspace{-0.5cm}
\end{figure*}

Figure \ref{fig:datastore1} exposes the meta-model used to structure the knowledge base KB. The entities refer to security properties and the relations encode associations among them. The entities in white are used to generated ADTrees, while those in grey are specialised for testing. The meta-model firstly associates attacks, techniques, security principles and security patterns. This is the result of  observations we made from the literature and some security documents, e.g., the CAPEC base or security pattern catalogues \cite{patrepo,Yskout2015}: we consider that an attack can be documented with more concrete attacks, which can be segmented into ordered steps; an attack step provides information about the target or puts an application into a state,  which are reused by a potential next step. Attack steps are performed with techniques and can be prevented with countermeasures. Security patterns are characterised with strong points, which are pattern features extractable from their descriptions. The meta-model also captures the inter-pattern relationships defined in  \cite{yskout2006system}, e.g.,  "depend" or "conflict". Countermeasures and strong points refer to the same notion of attack prevention. But finding direct relations between countermeasures and strong points is  tedious as these properties have different purposes. To solve this issue, we used a text mining and a clustering technique to group the countermeasures that refer to the same security principles, which are desirable security properties. To link clusters and strong points, we chose to focus on these security principles as mediators. We organised security principles into a hierarchy, from the most abstract to the most concrete principles. We provide a complete description of this hierarchy in \cite{SR17b}. In short, we collected and organised 66 security principles covering the security patterns of the catalogue given in \cite{Yskout2015}. The hierarchy has four levels, the first one being composed of elements labelled by the most abstract principles, e.g., \enquote{Access Control}, and the lower level exhibiting the most concrete principles, e.g., \enquote{File Authorization}.

Furthermore,  every attack step is associated to one test case structured with the Given When Then (GWT) pattern. We indeed consider in this paper that a test case is a piece of  code that lists stimuli supplied to an AUT and responses checked by assertions assigning (local) verdicts. To make test cases readable and re-usable, we use the behaviour driven approach using the pattern \enquote{Given When Then} (shortened GWT) to break up test cases into several sections: 

\begin{itemize}
	\item Given sections aim at putting the $AUT$ into a known state;
	\item When sections trigger some actions (stimuli);
	\item Then sections are used to check whether the conditions of success of the test case are met with assertions. In the paper, the Then sections are used to check whether an $AUT$ is vulnerable to an attack step $st$. In this case, the Then section returns the verdict \enquote{$Pass_{st}$}. Otherwise, it provides the verdict \enquote{$Fail_{st}$}. When a unexpected event occurs, we also assume that \enquote{$Inconclusive_{st}$} may be returned. \end{itemize}

The meta-model of Figure \ref{fig:datastore1} associates an attack step with a GWT test case by adding three entities (Given When and Then section) and relations. In addition, a test case section is linked to one procedure, which implements it. A section or a procedure can be reused with several attack steps or security patterns. %We indeed quickly observed that some attacks share preconditions, actions, or outcomes.
The meta-model also reflects the fact that an attack step is associated with one \enquote{Test architecture} and with one \enquote{Application context}. The former refers to textual paragraphs explaining the points of observation and control, testers or tools required to execute the attack step on an $AUT$. An application context refers to a family, e.g., Android applications, or Web sites. As a consequence, a GWT test case section (and procedure) is classified according to one application context and one attack step or pattern consequence.

We finally updated the meta-model in such a way that a security pattern is also associated to generic UML sequence diagrams, themselves arranged in Application contexts. Security pattern catalogues often provide  UML sequence diagrams expressing the security pattern behaviours or structures. These diagrams often help correctly implement a security pattern with regard to an application context.

\subsection{Data Integration}

We integrated data into KB by collecting them from heterogeneous sources: the CAPEC base, several papers dealing with security principles \cite{saltzer1975protection,viega2001building,Scambray2003,dialani2002transparent, meier2006web}, the pattern catalogue given in \cite{Yskoutcatalog} and the inter-pattern relations given in \cite{yskout2006system}. We details the data acquisition and integration steps in \cite{SR17a}. Six manual or automatic steps are required: Steps 1 to 5 give birth to databases that store security properties and establishing the different relations presented in Figure \ref{fig:datastore1}. Step 6 consolidates them so that every entity of the meta-model is related to the other ones as expected. The steps 1,2 and 6  are automatically done with tools. 

The current knowledge base KB includes information about  215 attacks (209 attack steps, 448 techniques), 26 security patterns, 66 security principles. We also generated 627 GWT test case sections (Given, When and Then sections) and 209 procedures. The latter are composed of comments explaining: which techniques can be used to execute an attack step and which observations reveal that the application is vulnerable. We manually completed 32 procedures, which cover 43 attack steps. %We used the testing framework Selenium and the penetration testing tool ZAProxy\footnote{https://www.owasp.org/index.php/OWASP\_Zed\_Attack\_Proxy\_Project}, which covers varied Web vulnerabilities. 
Security patterns are associated to at least one UML diagram. This knowledge base is available in \cite{data}.

It is worth noting that KB can be semi-automatically updated if new security data are available. If a new threat or type of attack is discovered and added to the CAPEC base, the steps 1, 2 and 5 have to be followed again. Likewise, if a new security pattern is proposed in the literature, the steps 3,4 and 5 have to be reapplied.

\section{Security Testing and Security Pattern Verification}
\label{sec:app}

\subsection{Approach Overview}
\label{sec:adtgen}

\begin{figure*}
	\centering
	\includegraphics[width=0.8\linewidth]{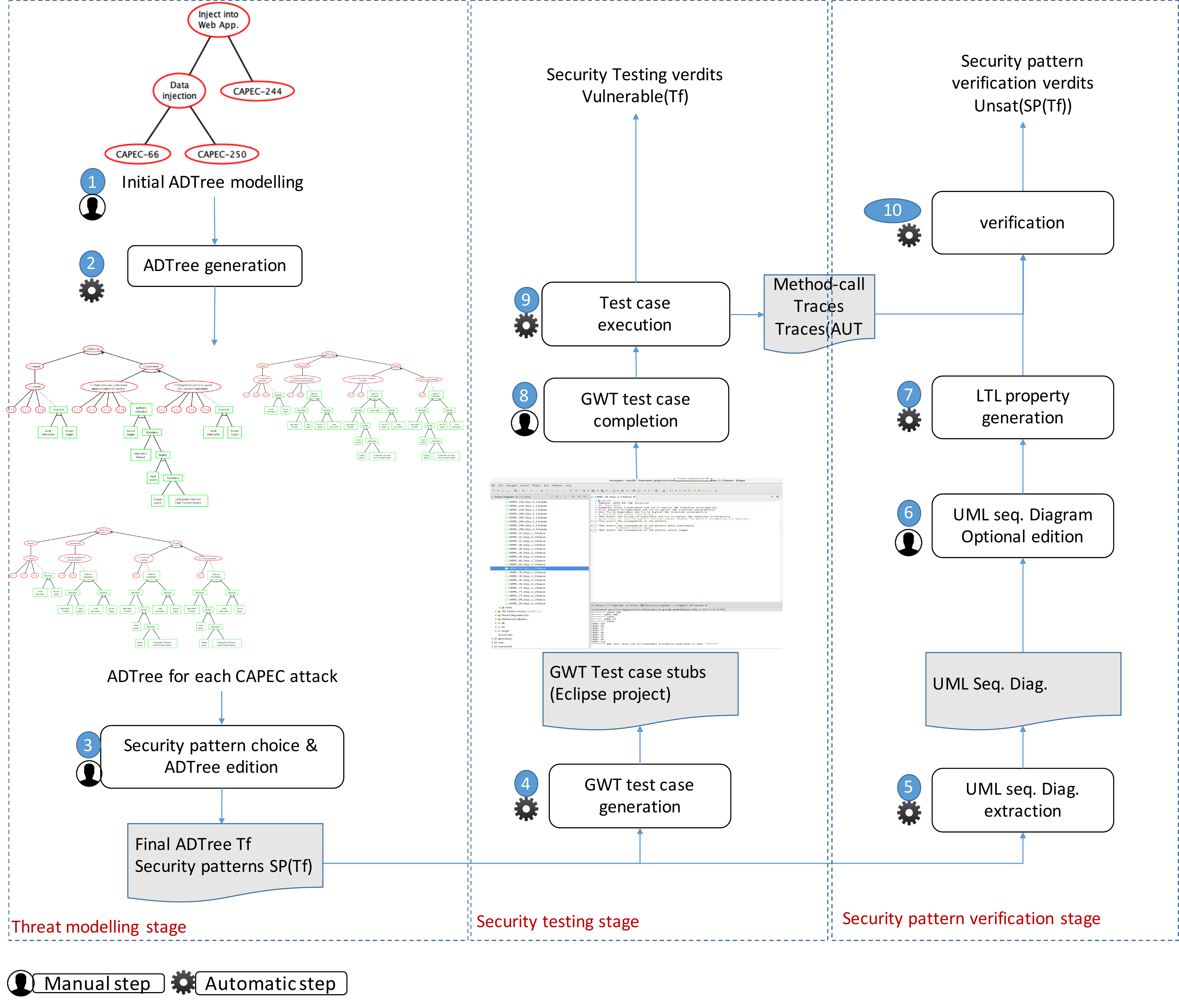}
	\caption{Overview of the 10 steps of the approach}
	\label{fig:overview}
	%\vspace{-0.5cm}
\end{figure*}

We present in this section our testing approach whose steps are illustrated in Figure \ref{fig:overview}. As illustrated in the figure, the purpose of this approach is threefold:

\begin{enumerate}
	\item \textbf{Threat modelling:} it firstly aims at guiding developers through the elaboration of a threat model (left side of the figure). The developer gives an initial ADTree expressing attacker capabilities (Step 1). By means of KB, this tree is automatically detailed  and completed with security patterns combinations expressing security solutions that may be put in place in the application design (Step 2). The tree may be modified to match the developer wishes (Step 3). The resulting ADTree, which is denoted $T_f$, captures possible attack scenarios and countermeasures given under the form of security pattern combinations. The set of security patterns chosen by the developer  is denoted $SP(T_f)$.

	\item \textbf{Security testing:} from $T_f$, the approach generates test case stubs, which are structured with the GWT pattern (Step 4). These stubs guide developers in the writing of concrete test cases (Step 8). The final test suite is executed on the $AUT$ to check whether the AUT is vulnerable to the attack scenarios expressed in the ADTree $T_f$ (Step 9). 
	
	\item \textbf{Security pattern verification:} the last part of the approach is devoted to checking whether security pattern behaviours hold in the $AUT$ traces. A set of generic UML sequence diagrams are extracted, from KB, for every security pattern in $SP(T_f)$ (Step 5). These show how security patterns classes or components should behave and help developers implement them in the application. These diagrams are usually adapted to match the application context (Step 6). The approach skims the UML sequence diagrams and automatically generates LTL properties encoding behavioural properties of the security patterns (Step 7). While the test case execution, the approach collects the $AUT$ method-call traces on which it checks whether the LTL properties are satisfied (Step 10).
	
\end{enumerate}

The remaining of this section describes more formally the steps depicted in Figure \ref{fig:overview}.

\subsection{Threat Modelling, Security Pattern Choice (Step 1 to 3)}
%\vspace{-0.5cm}

\noindent \textbf{Step 1: Initial ADTree Design}

The developer draws a first ADTree $T$ whose root node represents some attacker's goals. This node may be refined with several layers of children to refine these goals. Different methods can be followed, e.g., DREAD \cite{owasp}, to build this threat model. We here assume that the leaves of this ADTree are exclusively labelled by CAPEC attack identifiers, since our knowledge base KB is framed upon the CAPEC base. %Otherwise, a semantic alignment may be required to replace some attack labels by similar attack identifiers available in the CAPEC base. 
Figure \ref{fig:injection} illustrates an example ADTree achieved for this step. The leaves of this tree are labelled by CAPEC attacks related to different kinds of injection-based attacks. Its describes in general terms attacker goals, but this model is not sufficiently detailed to generate test cases or to choose security solutions.

\begin{comment}[htbp]
\centering
%\scalebox{0.7}[0.5]{
\includegraphics[width=0.4\linewidth]{figures/Injection} %}
\caption{ADTree example}
\label{fig:Injection}
\end{comment}

\noindent \textbf{Step 2: ADTree Generation}
\begin{figure}[htbp]
\centering
%\scalebox{0.7}[0.5]{
\includegraphics[width=0.7\linewidth]{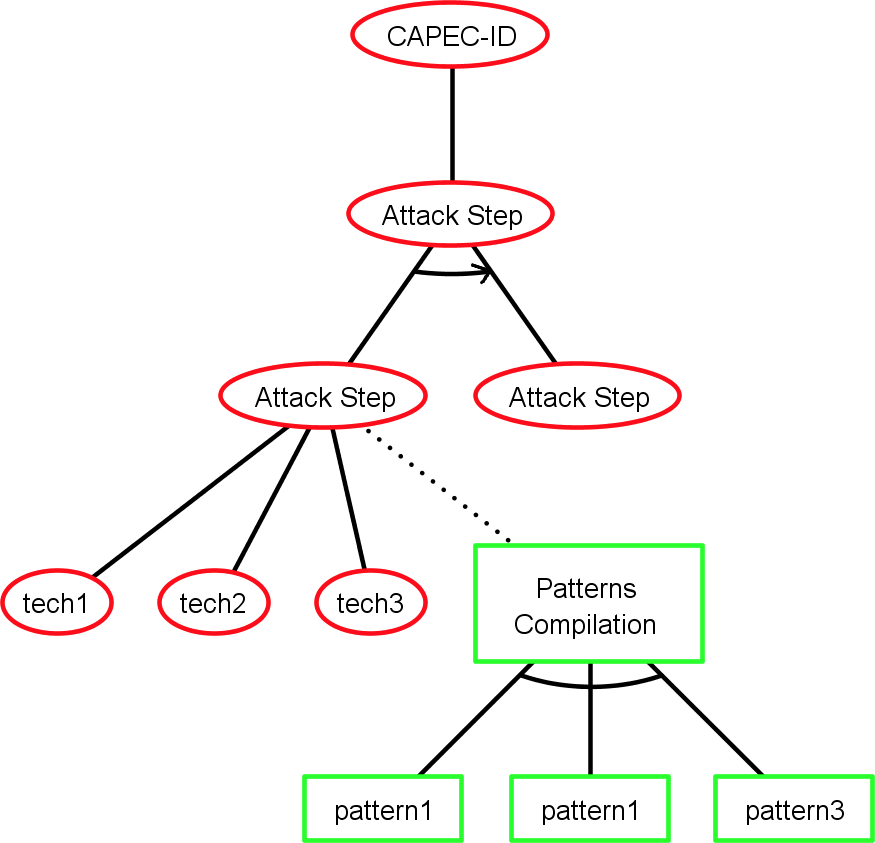} %}
\caption{Genenal form of the generated ADTrees}
\label{fig:genADTree}
\end{figure}

\begin{figure*}
	\centering
	\scalebox {1}[.9]{
		\includegraphics[width=.7\linewidth]{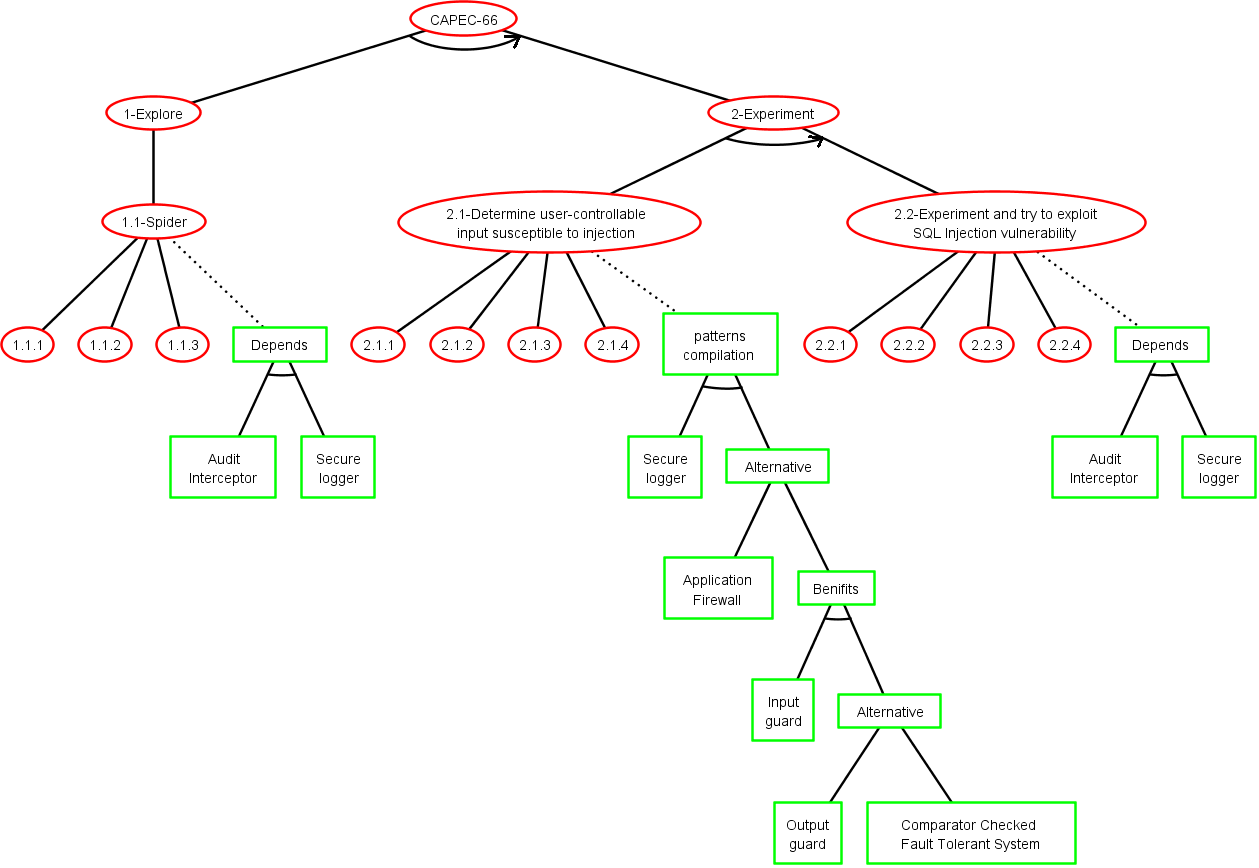}}
	\caption{ADTree of the Attack CAPEC-66}
	\label{fig:adtree66}
	%\vspace{-.5cm}
\end{figure*}

\begin{figure*}
	\centering
	\scalebox {1}[1]{
		\includegraphics[width=1\linewidth]{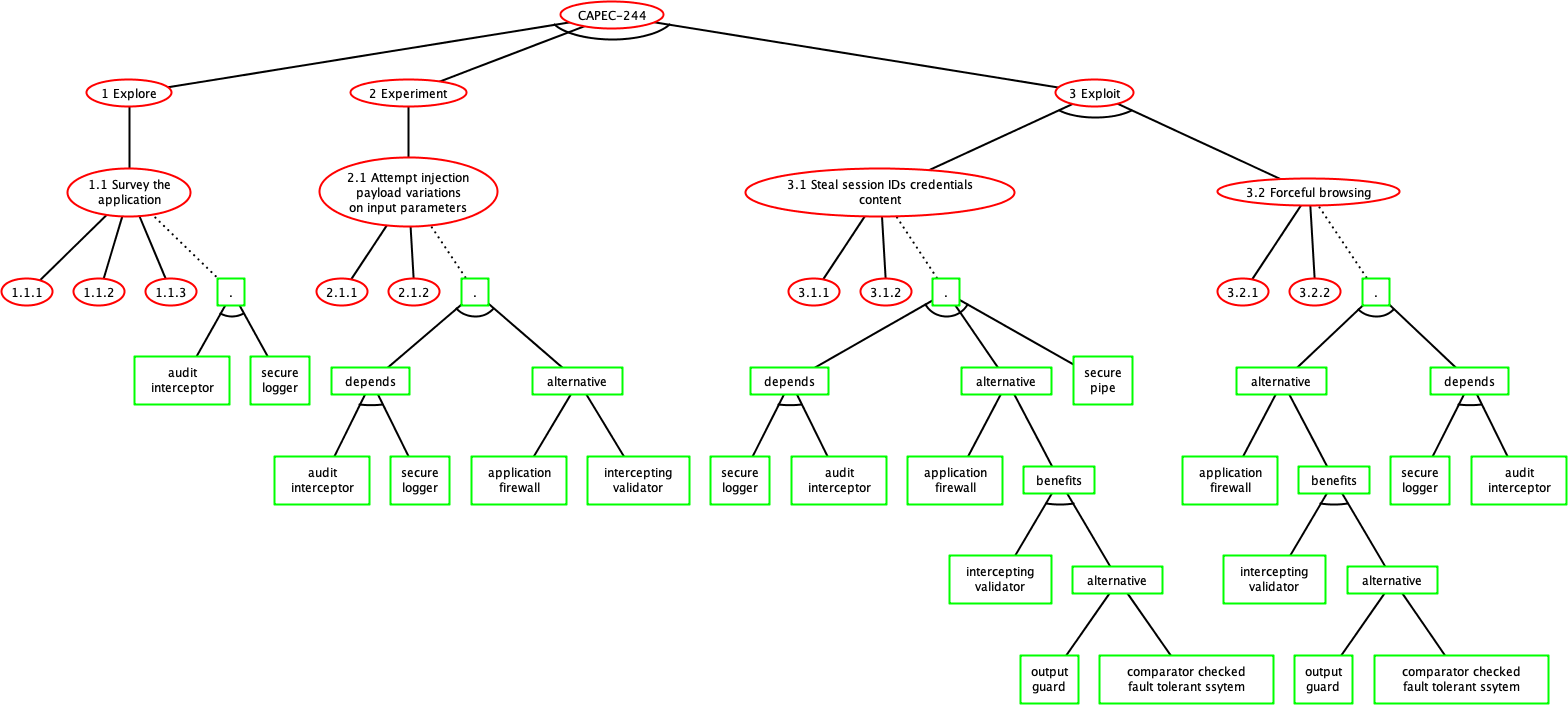}}
	\caption{ADTree of the Attack CAPEC-244}
	\label{fig:adtree244}
	%\vspace{-.5cm}
\end{figure*}

\begin{comment}[htbp]
\centering
%\scalebox{0.7}[0.5]{
\includegraphics[width=0.4\linewidth]{figures/xor} %}
\caption{Conflicting pattern representation with ADTree}
\label{fig:genADTree2}
\end{comment}

\begin{comment}[htbp]
\centering
\subfigure[Generic example of ADTree]{
\includegraphics[width=0.6\linewidth]{figure/generictree}}\quad
\subfigure[Conflicting pattern representation with ADTree]{		\includegraphics[width=0.4\linewidth]{figure/xor}}\\
\caption{}
\label{fig:genADTree}
\end{comment}

KB is now queried to complete $T$ with more details about the attack execution phase and with defense nodes labelled by security patterns. For every leave of $T$ labelled by an attack $A$, an ADTree $T(A)$, is generated from KB. %For the test case generation, we want that $T(A)$ includes sequences of attack steps that have to be executed in the right order to perform the attack. It also has to include defense nodes expressing combinations of security patterns whose purposes are to prevent the attack steps. 
We refer to \cite{SR17b} for the description  of the ADTree generation. 

We have implemented the ADTree generation with a tool, which takes attacks of KB and yields XML files. These can be edited with the tool \textit{ADTool} \cite{kordy2012attack}. For instance, Figures \ref{fig:adtree66} and \ref{fig:adtree244} show the ADTrees generated for the attacks CAPEC-66 and CAPEC-244. The ADTrees generated by this step are composed of several levels of attacks, having different levels of abstraction. The attack steps have child nodes referring to attack techniques, which indicate how to carry out the step. For instance the technique 1.1.1 is \enquote{Use a spidering tool to follow and record all links and analyze the web pages to find entry points. Make special note of any links that include parameters in the URL}. An attack step node is also linked to a defense node expressing security pattern combinations. Some nodes express inter-pattern relations. For instance, the node labelled by \enquote{Alternative} has children expressing several possible patterns to counter the attack step.

%In short, this ADTree is structured by the meta-model of Figure \ref{fig:datastore1} and 

Figures \ref{fig:adtree66} and \ref{fig:adtree244} also reveal that our generated ADTrees follow the structure of our meta-model of Figure \ref{fig:datastore1}. This structure has the generic form given in Figure \ref{fig:genADTree}: ADTrees have a root attack node, which may be disjunctively refined with other attacks and so forth. The most concrete attack nodes are linked to defense nodes labelled by security patterns. We formulate in the next proposition that these nodes or sub-trees also are encoded with specific ADTerms, which shall be used for the test case generation:

\begin{proposition}
	An ADTree $T(A)$ achieved by the previous steps has an ADTerm  $\iota(T(A))$ having one of these forms:
	\begin{enumerate}
		\item $\vee^p (t_1, \dots, t_n)$ with $t_i (1 \leq i \leq n)$ an ADTerm also having one of these forms:
		
		\item $\overrightarrow{\wedge}^p(t_1,\dots, t_n)$ with $t_i (1 \leq i \leq n)$ an ADTerm having the form given in 2) or 3);

		% \overrightarrow{\wedge}^p(t_1,\dots t_n), with $t_i$ an ADTerm also having this form or,
		%	\wedge^o(sp_1,\dots, sp_m)
		\item $c^p(st,sp)$, with $st$ an ADTerm expressing an attack step and $sp$ an ADTerm modelling a  security pattern combination. 
	\end{enumerate}
\end{proposition}

The first ADTerm expresses child nodes labelled by more concrete attacks. The second one represents sequences of attack steps. The last ADTerm is composed of an attack step $st$ refined with techniques, which can be counteracted by a security pattern combination $sp=\wedge^o(sp_1,\dots, sp_m)$. In the remainder of the paper, we denote the last expression $c^p(st,sp)$ a \textit{Basic Attack Defence Step}, shortened as BADStep:

\begin{definition}[Basic Attack Defence Step (BADStep)]
A BADStep $b$ is an ADTerm of the form $c^p(st, sp)$, where $st$ is a step only refined with techniques and $sp$ an ADTerm of the form:
	\begin{enumerate}
		\item $sp_1$, with $sp_1$ a security pattern,
		
		\item $\wedge^o(sp_1,\dots,sp_m)$ modelling the conjunction of the security patterns $sp_1,\dots,\\sp_m (m>1)$.
	\end{enumerate}
	$\defense(b)=_{def} \{sp_1 \}$ iff $sp=sp_1$, or  $\defense(b)=_{def} \{sp_1,\dots,sp_m \}$ iff $sp=\wedge^o(sp_1,\dots,sp_m)$.\\
	$\BADStep(T)$ denotes the set of BADSteps of the ADTree $T$. 
\end{definition}

\noindent \textbf{Step 3: Security Pattern Choice and ADTree Edition}
\begin{figure*}
	\centering
	\scalebox {1}[1]{
		\includegraphics[width=1\linewidth]{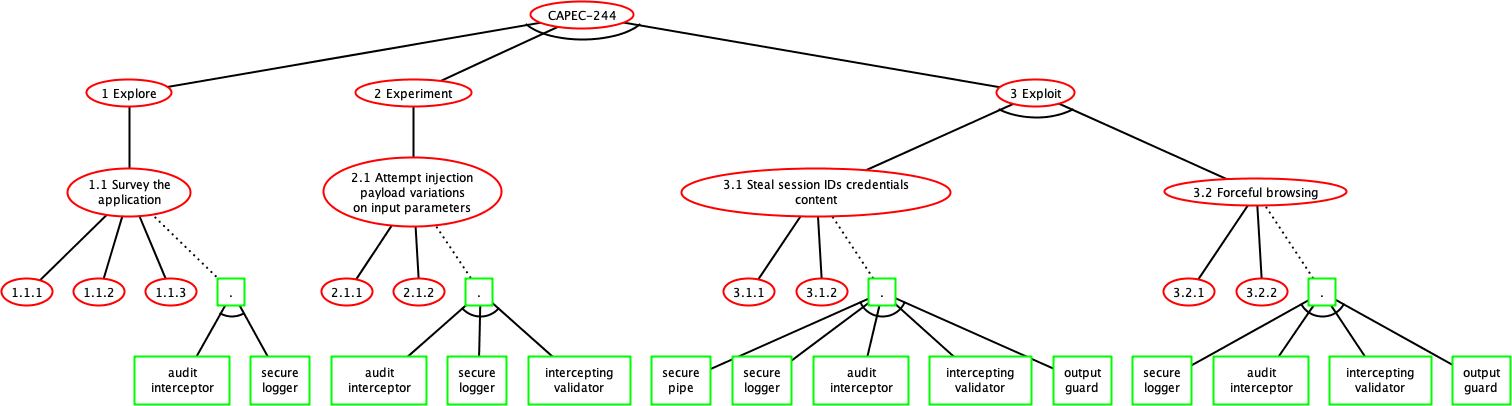}}
	\caption{Final ADTree of the Attack CAPEC-244}
	\label{fig:adtree244-2}
	%\vspace{-.5cm}
\end{figure*}

The developer may now edit every ADTree $T(A)$ generated by the previous step and choose security patterns when several possibilities are available. We assume that
the defense nodes linked to attack nodes have conjunctive refinements of nodes labelled by security patterns only. Figure \ref{fig:adtree244-2} depicts an example of modified ADTree of the attack CAPEC-244. 

Every attack node $A$ of the initial ADTree $T$ is now automatically replaced with the ADTree $T(A)$. This step is achieved by substituting every term $A$ in the ADTerm $\iota(T$) by $\iota(T(A))$. We denote $\iota(T_f)$ the resulting ADTerm and $T_f$ the final ADTree. It depicts a logical breakdown of the options available to an attacker and the defences, materialised with security patterns, which have to be inserted into the application model and then implemented. The security pattern set found in $T_f$ is denoted $SP(T_f)$.

%non necessaire  recursively transform $c^p(f(A_1,\dots A_n),SP)$ with f in \{\overrightarrow{\wedge}^p, \vee^p, \wedge^p\}-> $f(c^p((A_1,SP), \dots c^p(A_n),SP)$ to place SP on the leaves of the tree
%Before initiating the GWT test case stub generation, we define some notations:
This step finally builds a report by extracting from KB the test architecture descriptions needed for executing the attacks on the $AUT$ and observing its reactions.

\subsection{Security Testing}
\label{sec:test}

We now extract attack-defense scenarios to later build test suites that will check whether attacks are effective on the $AUT$. An attack-defense scenario is a minimal combination of events leading to the root attack, minimal in the sense that, if any event of the attack-defense scenario is omitted, then the root goal will not be achieved. 

%Informally speaking, as we do not unfold the concrete attacks (which are sequential conjunctions of BADSteps), of an ADTrees $T_f$, the ADTerm $\iota(Tf)$ only includes conjunctive and disjunctive operators. 

The set of attack-defense scenarios of $T_f$ are extracted by means of the disjunctive decomposition of $\iota(Tf)$: 

\begin{definition}[Attack scenarios]
	Let $T_f$ be an ADTree and $\iota(T_f)$ be its ADTerm. The set of Attack scenarios of $T_f$, denoted $SC(T_f)$ is the set of clauses of the disjunctive normal form of $\iota(T_f)$ over $BADStep(T_f)$. 
	
	$BADStep(s)$ denotes the set of BADSteps of a scenario $s$.
\end{definition}

An attack scenario $s$ is still an ADTerm. Its satisfiability means that the main goal of the ADTree $T_f$ is feasible by achieving the scenario formulated by $s$. $BADStep(s)$ denotes the set of BADSteps of $s$.\\ %We also denote $SP(s)$ the security pattern set found in $s$: $SP(s)=\{sp  \mid \exists b\in BADStep(s): sp \in \defense(b) \}$. By extension, $SP(T_f)$ is the security pattern set of $\iota(T_f)$, found in all its scenarios.\\ %The set $BADStep(s)$ shall be used to build test suites, as explained in the following step.

%We also denote $SP(s)$ and $SP(T_f)$ the security pattern sets found in an attack-defense scenario $s$ and in $\iota(T_f)$. %$SP(T_f)=\{sp  \mid \exists s \in b\in Sc(T_f), b \in \BADStep(s): sp \in \defense(b) \}$. 

\noindent \textbf{Step 4: Test Suite Generation}

Let $s \in SC(Tf)$ be an attack-defense scenario and $b=c^p(st, sp) \in BADStep(s)$ a BADSteps of $s$. Step 4 generates the GWT test case $TC(b)$ composed of 3 sections extracted from KB  with the relations $testG$, $testW$ and $testT$: we have one Given section, one When section and one Then section, each related to one procedure. This Then section aims to assert whether the $AUT$ is vulnerable to the attack step $st$ executed by the When section.
	
The final test suite $TS$, derived from an ADTree $T_f$,  is obtained after having iteratively applied this test case construction on the scenarios of $SC(T_f)$. This is captured by the following definition:

\begin{definition}[Test suites]
	Let $T_f$ be an ADTree, $s \in SC(Tf)$ and $b \in BADStep(s)$.\\
	$TS= \{TC(b) \mid b=c^p(st, sp) \in BADStep(s) \text{ and } s \in SC(T_f) \}$.
	
\end{definition}

\begin{figure}
	\begin{scriptsize}
		\begin{lstlisting}
	@capec244
	Feature: CAPEC-244: Cross-Site Scripting via Encoded URI Schemes 
	#1. Explore
	Scenario: Step1.1 Survey the application
	Given prepare to Survey the application
	When Try to Survey the application
	# assertion for attack step success
	Then Assert the success of Survey the application
		\end{lstlisting}
	\end{scriptsize}
	\caption{The test case stub of the first step of the attack CAPEC 244}
	\label{fig:feature}
	%\vspace{-0.2cm}
\end{figure}

\begin{figure}
	\begin{lstlisting}
	@When("Try to Survey the application for user -controllable inputs")
	public void trysurvey(){
	//  Try one of the following techniques :
	//1.  Use a spidering tool to follow and record all links and analyze the web pages to find entry points. Make special note of any links that include parameters in the URL.
	//2.  Use a proxy tool to record all user input entry points visited during a manual traversal of the web application.
	//3.  Use a browser to manually explore the website and analyze how it is constructed. Many browsers' plugins are available to facilitate the analysis or automate the discovery.
	String url ="";
	ZAProxyScanner j = new ZAProxyScanner("localhost", 8080, "zap");
	j.spider(url);
	}
	@Then("Assert the success of Survey the application for user-controllable inputs")
	public void asssurvey(){
	//  Assert one of the following indications :
	//  -A list of URLs, with their corresponding parameters (POST, GET, COOKIE, etc.) is created by the attacker.
	ZAProxyScanner j = new ZAProxyScanner("localhost", 8080, "zap");
	int x = j.getSpiderResults(j.getLastSpiderScanId())
	.size();
	Assert.assertTrue(x>0);
	}}\end{lstlisting}
	\caption{The procedure related to the When and Then sections of Figure \ref{fig:feature}}
	\label{fig:proc}
	\vspace{-0.5cm}
\end{figure}

We have implemented these steps to yield GWT test case stubs compatible with the Cucumber framework \cite{cucumber}, which supports a large number of languages. Figure \ref{fig:feature}  gives a test case stub example obtained with our tool from the first step of the attack CAPEC-244 depicted in Figure \ref{fig:adtree244}. The test case lists the Given When Then sections in a readable manner. Every section is associated to a generic procedure stored into another file. The procedure related to the When and Then sections are given in Figure \ref{fig:proc}. The comments come from KB and  the CAPEC base. In this example, the procedure includes a generic block of code, which may be reused with several applications; the \enquote{getSpider()} method relates to the call of the ZAProxy\footnote{https://www.owasp.org/index.php/OWASP\_Zed\_Attack\_Proxy\_Project} tool, which crawls a Web application to get its URLs.\\

\noindent \textbf{Step 8: Test Case Stub Completion}

In the beginning of this step, the test case procedures are generic, which means that they are composed of comments or generic block of codes that  help developers complete them. In the previous test case example, it only remains for the developer to write the initial URL of the Web application before testing whether it can be explored. Unfortunately, with other test cases, the developer might have to implement it completely.

After this step, we assume that the test cases are correctly developed with assertions in Then sections as stated in Section \ref{sec:datastore}: a Then section of a test case $TC(b)$ returns the verdict "$Pass_{st}$" if an attack step $st$ has been successfully applied on the $AUT$ and "$Fail_{st}$" otherwise; when $TC(b)$ returns an unexpected exception or fault, we get the verdict "$Inconclusive_{st}$".\\ 

\noindent \textbf{Step 9: Test Case Execution}

The experimentation of the $AUT$ with the test suite $TS$ is carried out in this step. A test case $TC(b)$ of $TS$, which aims at testing whether the $AUT$ is vulnerable to an attack step $st$ leads to a local verdict denoted $\Verdict(TC(b) \vert \vert AUT)$: 

\begin{definition}[Local Test Verdicts]
 	Let $AUT$ be an application under test, $b=c^p(st,sp) \in BADStep(T_f)$, and $TC(b)\in TS$ be a test case.\\ 
 	$\Verdict(TC(b) \vert \vert AUT)=$
 		\begin{itemize}
 			\item $Pass_{st}$, which means $AUT$ is  vulnerable to the attack step $st$;
 		\item $Fail_{st}$, which means $AUT$ does not appear to be vulnerable to the attack step $st$;
 		
 		\item $Inconclusive_{st}$, which means that various problems occurred while the test case execution.
 		
 		%\item $\Unsat^b(sp_1)=_{def} true$ if $\exists p \in P(sp_1), \exists t \in Traces(AUT), t \nvDash p$; otherwise, $\Unsat^b(sp_1)$ $=_{def}false$; 
 	\end{itemize}
 \end{definition}

We finally define the final verdicts of the security testing stage with regard to the ADTree $T_f$. These verdicts are given with the predicates $\Vulnerable(T_f)$ and $\Inconclusive(T_f)$ returning boolean values. The intermediate predicate $\Vulnerable(b)$ is also defined on a BADStep $b$ to evaluate a substitution $\sigma : BADStep(s) \rightarrow \{true,false\}$ on an attack-defense scenario $s$. %This substitution is used to evaluate attack-defense scenarios. 
A scenario $s$ holds if the evaluation of the substitution $\sigma$ to $s$, i.e., replacing every BADStep term $b$ with the evaluation of  $\Vulnerable(b)$, returns true. 
The predicate  $\Vulnerable(s)$ expresses whether an attack-defense scenario of $T_f$ holds. In that case, the threat modelled by $T_f$ can be achieved on $AUT$. This is defined with the predicate $\Vulnerable(T_f)$: 

\begin{definition}[Security Testing Verdicts]
	Let $AUT$ be an application under test, $T_f$ be an ADTree, $s \in SC(T_f)$ and $b=c^p(st,sp) \in BADStep(s)$.  
	
	\begin{enumerate}
		\item $\Vulnerable(b)=_{def} true$  if $\Verdict(TC(b) \vert \vert AUT)= Pass_{st}$; otherwise,  $\Vulnerable(b)=_{def} false$;
		
		\item $\Vulnerable(s)=_{def} true$ if $eval(s\sigma)$ returns true, with  $\sigma:BADStep(s)\rightarrow \{true,false\}$ the substitution $\{b_1 \rightarrow \Vulnerable(b_1), \dots, b_n \rightarrow \Vulnerable(b_n) \}$; otherwise, $\Vulnerable(s)=_{def} false$;
		
		\item $\Inconclusive(s)=_{def}true$ if $\exists b \in \BADStep(s)$:
		$\Verdict(TC(b) \vert \vert$ $AUT)=Inconclusive_{st}$; otherwise, $\Inconclusive(s)=_{def}false$.

		\item $\Vulnerable(T_f)=_{def} true$ if $\exists s \in SC(T_f): \Vulnerable(s)=true$; otherwise, $\Vulnerable($ $T_f)=_{def} false$;
		
		\item $\Inconclusive(T_f)=_{def}true$ if $\exists s \in SC(T_f),$ $\Inconclusive(s)=true$; otherwise, $\Inconclusive(T_f)$ $=_{def}false$.
		
		%\item $\Unsat^b(SP(T_f))=_{def}true$ if $\exists sp \in SP(T_f), \Unsat^b(sp)= true$; otherwise, $\Unsat^c(\\SP(T_f))$ $=_{def}false$; %if $\forall b \in \BADStep(T_f)$,  $\Verdict(TC(b) \vert \vert IUT) \in VUL, NVUL$	
	\end{enumerate}
\end{definition}

\subsection{Security Pattern Verification}

Our approach also aims at checking whether security patterns are correctly implemented in the application. The security testing stage is indeed insufficient because the non-detection of vulnerability in the $AUT$ does not imply that a security pattern is correctly implemented. As stated earlier, we propose to generate LTL properties that express the behavioural properties of a security pattern. Then, these are used  to check whether they hold on the $AUT$ traces. The originality of our approach resides in the fact that we do not ask developers for writing formal properties, we propose to generate them by means of KB.\\

\noindent \textbf{Steps 5 and 6: UML Sequence Diagram Extraction and Modification}

After the threat modelling stage, this step starts by extracting from KB  a list of generic UML sequence diagrams for each security pattern in $SP(T_f)$.  These diagrams show how a security pattern should behave once it is correctly implemented, i.e., how objects interact in time. We now suppose that the developer implements every security pattern in the application. At the same time, he/she may adapt the behaviours illustrated in the UML sequence diagrams. In this case, we assume that the diagrams are updated accordingly.

\begin{figure}
	\centering
	\includegraphics[width=1\linewidth]{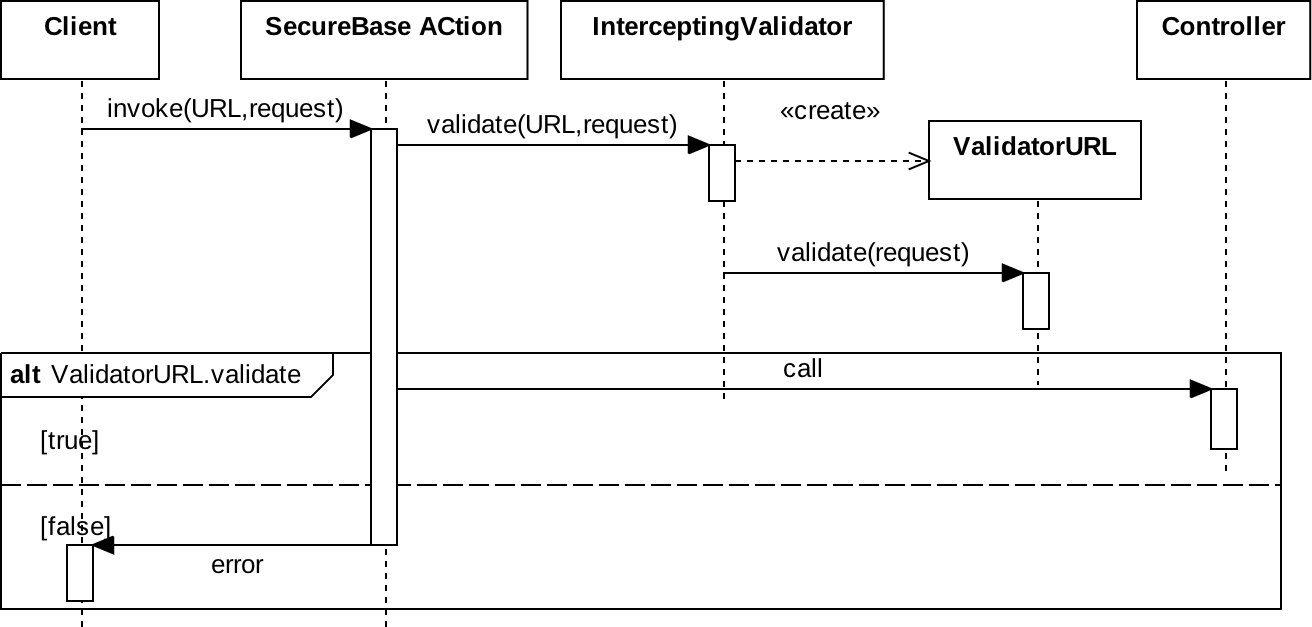}
	\caption{UML sequence diag. of the pattern Intercepting Validator}
	\label{fig:case}
	%\vspace{-.5cm}
\end{figure}

Figure \ref{fig:case} illustrates an example of UML sequence diagram for the security pattern \enquote{Intercepting Validator}. The diagram shows the interactions between an external Client, the pattern and the application, but also the interactions among the objects of the pattern. %The pattern classes may be adapted: for instance, if the name of the method \enquote{process} has to be modified by \enquote{send}, the new label must be of the form \enquote{process/send} to express the substitution.  
Here, the Intercepting Validator Object is called to validate requests. These are given to another object ValidatorURL, which filters the request with regard to the URL type. If the request is valid, it is processed by the application (Controller object), otherwise an error is returned to the client side.\\

\noindent \textbf{Step 7: Security Pattern LTL Property Generation}

This step automatically generates LTL properties from UML sequence diagrams by detecting the  cause-effect relations among method calls and expressing them in LTL. Initially, we took inspiration in the method of Muram et al. \cite{MuramTZ14}, which transforms activity diagrams into LTL properties. Unfortunately, security patterns are not described with activity diagrams, but with sequence diagrams. This is why we devised 20 conversion schemas  allowing to transform UML sequence diagram constructs, composed of two or three successive actions, into UML activity diagrams. Table \ref{tab:rules} gives 6 of these schemas. Intuitively, these translate two consecutive method calls found in a sequence diagram by activity diagrams composed of action states. The other schemas (not all given in Table \ref{tab:rules}) are the results of slight adaptations of the five first ones, where the number of objects or the guards have been modified. For instance, the last schema of Table \ref{tab:rules} is an adaptation of the first one, which depicts interactions between two objects instead of three. 
 
Then, we propose 20 rules to translate these activity diagrams into LTL properties. The last column of Table \ref{tab:rules} lists 6 of these rules. Some of these rules are based on those proposed by Muram et al, but we devised other rules related to our own activity diagrams, which are more detailed. For instance, we take into account the condition state in the second rule to produce more precise LTL properties. 

%They proposed 5 rules that translate 5 activity constructs, composed of two consecutive actions, into LTL primitives. We modified some of these rules to make them more precise (in particular the conditional construct). 

At the end of this step, we consider having a set of LTL properties $P(sp)$ for every security pattern $sp \in SP(T_f)$. Although the LTL properties of $P(sp)$ do not necessarily cover all the possible behavioural properties of a security pattern $sp$, this process offers the advantages of not asking developers for writing LTL formula or to instantiate generic LTL properties to match the application model or code.

%, which is considered as a hard and error-prone task.
%d'autres patterns possibles

\begin{table}[htbp]
	\centering
	\caption{UML sequence diagrams to LTL properties transformation rules.}
	\label{tab:rules}
	%\begin{scriptsize}
		\begin{tabular}[t]{|m{3cm}|m{2.1cm}|m{2.3cm}|}
			\hline Sequence Diag.&Activity Diag.&LTL properties\\ 
			\hline \includegraphics[width=1\linewidth]{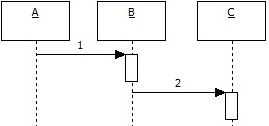} & \includegraphics[width=.3\linewidth]{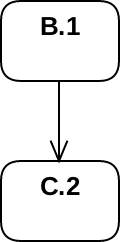} &$\square (B.1  \longrightarrow \lozenge C.2)$ \\ 
			\hline \includegraphics[width=1\linewidth]{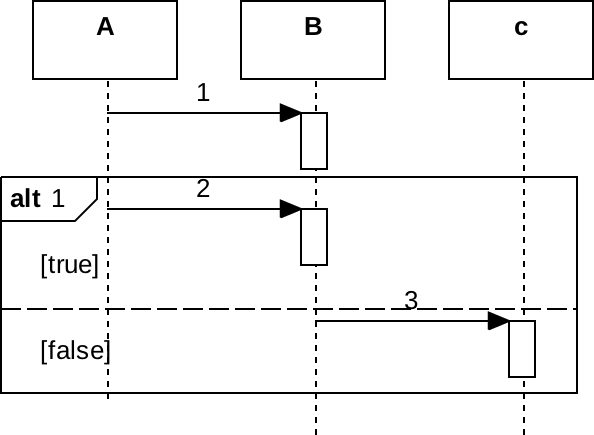} & \includegraphics[width=.8\linewidth]{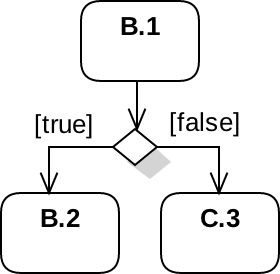} &$\square ( B.1 \longrightarrow \lozenge $ $B.2 )$ $xor$ $(\neg B.1$ $\longrightarrow  \lozenge C.3)) $\\ 
						 	%$\square (B.1  \longrightarrow (\lozenge B.2) xor(\lozenge C.3))$\\ 
			\hline \includegraphics[width=1\linewidth]{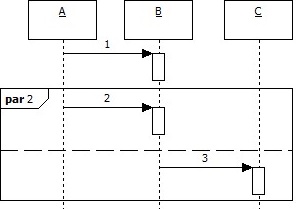} & \includegraphics[width=1\linewidth]{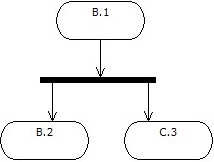} & $\square (B.1  \longrightarrow (\lozenge B.2) and (\lozenge C.3))$ \\ 
			\hline \includegraphics[width=1\linewidth]{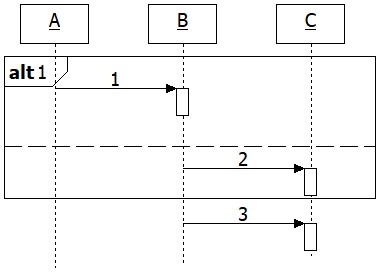} & \includegraphics[width=.8\linewidth]{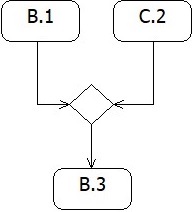} & $\square (B.1 xor C.3  \longrightarrow \lozenge B.3)$ \\ 
			\hline \includegraphics[width=1\linewidth]{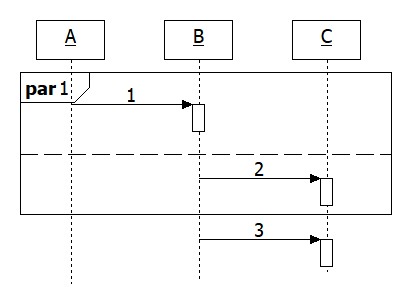} & \includegraphics[width=.8\linewidth]{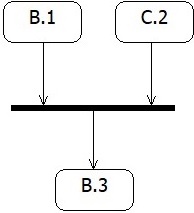} & $\square (B.1 and C.3  \longrightarrow \lozenge B.3)$ \\ 
			\hline \includegraphics[width=.7\linewidth]{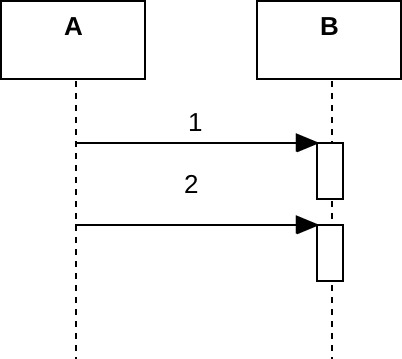} & \includegraphics[width=.3\linewidth]{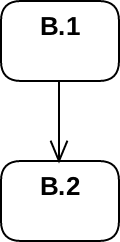} & $\square (B.1 \longrightarrow \lozenge B.2)$ \\ 
			\hline 
		\end{tabular} 
	%\end{scriptsize}
\end{table}

From the example of UML sequence diagram given in Figure \ref{fig:case}, 4 LTL properties are generated. Table \ref{prop} lists them. These capture the cause-effect relations of every pair of methods found in the UML sequence diagram.

\begin{comment}[!h]
\centering
\includegraphics[width=1\linewidth]{figure/act/case2}
\caption{Authentication Enforcer activity diagram}
\label{fig:case2}
\end{comment}

\begin{table}[h]
	\caption{LTL properties for the pattern Intercepting Validator.}
	\label{prop}
	\setlength\tabcolsep{1.5pt} % default value: 6pt
	
	\begin{tabular}{|p{0.3cm}|p{8.1cm}|}
		\hline
		
		$p_1$& $\square (SecureBaseAction.invokes$ $\longrightarrow\lozenge InterceptingValidator.validate$\\ 
		\hline 
		$p_2$& $\square (InterceptingValidator.validate  \longrightarrow \lozenge ValidatorURL.create)$\\ 
		\hline 
		$p_3$& $\square (ValidatorURL.create$  $\longrightarrow \lozenge ValidatorURL.validate)$\\ 
		\hline
		$p_4$& $  \square ( (ValidatorURL.validate \longrightarrow \lozenge Controller.call )$ $xor$ $(\neg ValidatorURL.validate \longrightarrow  \lozenge SecureBaseAction.error)) $\\ 
		
		\hline 
	\end{tabular} 
\end{table}

\noindent \textbf{Step 10: Security Pattern Verification}

As stated earlier, we consider that the $AUT$ is instrumented with a debugger or similar tool to collect the methods called in the application while the execution of the test cases of $TS$. After the test case execution, we hence have a set of method call traces denoted $Traces(AUT)$. 

A model-checking tool is now used to detect the non-satisfiability of LTL properties on $Traces(AUT)$. Given a security pattern $sp$, the predicate $Unsat^b(sp)$ formulates the non-satisfiability of a LTL property of $sp$ in $Traces(AUT)$. The final predicate $\Unsat^b(SP(T_f))$ expresses whether all the LTL properties of the security patterns given in $T_f$ hold. 

\begin{definition}[Security Pattern Verification Verdicts]
	Let $AUT$ be an application under test, $T_f$ be an ADTree, and $sp \in SP(T_f)$ be a security pattern.   
	
	\begin{enumerate}
	\item $\Unsat^b(sp)=_{def} true$ if $\exists p \in P(sp), \exists t \in Traces(AUT), t \nvDash p$; otherwise, $\Unsat^b(sp)$ $=_{def}false$; 
		
	\item $\Unsat^b(SP(T_f))=_{def}true$ if $\exists sp \in SP(T_f), \Unsat^b(sp)= true$; otherwise, $\Unsat^b(SP(T_f))$ $=_{def}false$; %if $\forall b \in \BADStep(T_f)$,  $\Verdict(TC(b) \vert \vert IUT) \in VUL, NVUL$	
\end{enumerate}
\end{definition}

Table \ref{tab:table1} informally summarises the meaning of some test verdicts and some corrections that may be followed in case of failure.

\begin{table}
	\begin{scriptsize}
		\begin{center}
			\caption{Test verdict Summary and Recommendations.} \label{tab:table1}
\setlength\tabcolsep{1.5pt} % default value: 6pt
\begin{tabular}{|p{.8cm}|p{1cm}|p{.7cm}|p{5.6cm}|}
	\hline
Vulnera-ble($T_f$)&$\Unsat^b($ $SP(T_f))$&Incon $(T_f)$&Corrective actions\\
				\hline
False&False&False&No issue detected\\ \hline
				True&False&False&At least one scenario is successfully applied on $AUT$. Fix the pattern implementation. Or the chosen patterns are inconvenient.\\ \hline 
				False&True&False&Some pattern behavioural properties do not hold. Check the pattern implementations with the UML seq. diag. Or another pattern conceals the behaviour of the former.\\ \hline
				True&True&False&The chosen security patterns are useless or incorrectly implemented. Review the ADTree, fix $AUT$.\\ \hline
				T/F&T/F&True&The test case execution crashed or returned unexpected exceptions. Check the Test architecture and the test case codes.\\ \hline
			\end{tabular}
		\end{center} 
	\end{scriptsize}
	
	\vspace{-0.5cm}
\end{table}

\section{Implementation}
\label{sec:impl}

Our approach is implemented in Java and is released as open source in \cite{data}. At the moment, the $AUT$ must be a Web application developed with any kind of language provided that the $AUT$ may be instrumented to collect method call traces. The prototype tool consists of three main parts. The first one comes down to a set of command lines allowing to build the knowledge base KB. The data integration is mostly performed by calling the tool Talend, which is specialised into the extract, transform, load (ETL) procedure. An example of knowledge base is available in \cite{data}. 

A second software program semi-automatically generates ADTrees and GWT test cases. ADTrees are stored into XML files, and may be edited with \textit{ADTool} \cite{kordy2012attack}. GWT test cases are written in Java with the Cucumber framework, which supports the GWT test case pattern. These test cases can be imported as an Eclipse project to be completed and executed. This software program also provides UML sequence diagrams stored in JSON files, which have to be modified to match the $AUT$ functioning. LTL properties are extracted from these UML sequence diagrams.

The last part of the tool is a tester that experiments Web applications with test cases and returns test verdicts. While the test case execution, we collect log files including method call traces. The LTL property verification on these traces is manually done by these steps: 1) the log files usually have to be manually filtered to remove necessary events 2) the tool Texada \cite{texada} is invoked to check the satisfiability of  every LTL property on the log files. This tool takes as inputs a log file, a LTL property composed of variables and a list of events specifying variables in the formula to be interpreted as a constant event. Texada returns the number of times that a  property holds in a log file. 
We have chosen the Texada tool as it offers good performance and can be used on large trace sets. But other tools could also be used, e.g., the LTL checker plugin of the ProM framework  \cite{Maggi}
or Eagle \cite{eagle}.

\section{Preliminary Evaluation}
\label{sec:eval}

First and foremost, it is worth noting that we carried out in \cite{SR17b} a first evaluation of the difficulty of using security patterns for designing secure applications. This evaluation was conducted on 24 participants and allowed us to conclude that the Threat modelling stage of our approach makes the security pattern choice and  the test case development easier and makes users more effective on security testing. In this paper, we propose another evaluation of the security testing and security pattern testing parts of our approach. This evaluation addresses the following research questions:

\begin{itemize}
	\item Q1: Can the generated test cases detect security issues? 
	\item Q2: Can the generated LTL properties detect incorrect implementation of patterns?  
	\item Q3: How long does it take to discover errors (Performance)?
	%How does our approach scale with the size of the test cases and of the traces?
\end{itemize}

%q1 abble to geenrate TC that detect security issues ? [resultats tests]
%q2 able to generate LTL prop. that detect incorrect impl. of patterns ? [result LTL prop. sur traces]
%Mq3 can tool be applied to real apps. ? [nb tests, nb LTL prop, temps]
%q4 does tool ease testin ? eval précédente .
%RQ2: Can our method detect incorrect applications of

\subsection{Empirical Setup}
We asked ten teams of two students to implement Web applications written in PHP as a part of their courses. They could choose to develop either a blog, or a todo list application or a RSS reader. Among the requirements, the Web applications had to manage several kinds of users (visitors, administrators, etc.), to be implemented in object-oriented programming, to use the PHP Data Objects (PDO) extension to prevent SQL injections, and to validate all the user inputs. As a solution to filter inputs, we proposed them to apply the security pattern Intercepting Validator. But its use was not mandatory.

Then, we applied our tool on these 10 Web applications in order to: 
\begin{itemize}
	\item test whether these are vulnerable to both SQL and XSS injections (attacks CAPEC-66 and CAPEC-244). With our tool, we generated the ADTrees of Figures \ref{fig:adtree66} and \ref{fig:adtree244},  along with  GWT test cases. We completed them to call the penetration testing tool ZAProxy (as illustrated in Figure \ref{fig:proc}). All the applications were vulnerable to the steps \enquote{Explore} of the ADTrees (application survey), therefore  we also experimented them with the test cases related to the steps \enquote{Experiment} (attempt SQL or XSS injections); 
	
	\item test whether the behaviours of the pattern Intercepting Validator are correctly implemented in the 10 Web applications. We took the UML sequence diagram of Figure \ref{fig:case} and adapted it ten times to match the context of every application. Most of the time, we had to change the class or method names, and to add as many validator classes as there are in the application codes. When a class or method of the pattern was not implemented, we leaved the generic name in the UML diagram. Then, we generated LTL properties to verify whether they hold in the application traces.
\end{itemize}

%The Web applications and results are available in \cite{data}.

\subsection{Q1: Can the generated test cases detect security issues?}

\subsubsection*{Procedure} 
to study Q1, we experimented the 10 applications with the 4 GWT test cases of the two Steps Explore and Experiment of the attacks CAPEC-66 and CAPEC-244. As these test cases call a penetration testing tool, which may report false positives, we manually checked the reported errors to only keep the real ones. We also inspected the application codes to examine the security flaw causes and to finally check whether the applications are vulnerable. 

\subsubsection*{Results} Table \ref{table:result1} provides the number of tests for both attacks (columns 2 and 3), the number of security errors detected by these tests (columns 4 and 5) and execution times in seconds (column 6). As a penetration testing tool is called, a large amount of malicious HTTP requests are sent to the applications in order to test them. The test number often depends on the application structure (e.g., number of classes, of called libraries, of URLs, etc.)  but also on the  number of forms available in an application.

Table \ref{table:result1} shows that errors are detected in half of the applications. After inspection, we observed that several inputs are not filtered in App. 1, 5 and 6. On the contrary, for App. 3 and 7 all the inputs are checked. However, the validation process is itself incorrectly performed or too straightforward. For example, in App. 3 the validation comes down to checking that the input exists, which is far from sufficient to block malicious code injections. For the other applications, we observed that they all include a correct validation process, which is called after every client request. After the code inspection and the testing process, we conclude that they seem to be protected against both XML and SQL injections. These experiments tend to confirm that our approach can be used to test the security of Web applications. 

\begin{table}[h]
	\caption{Results of the security testing stage: number of requests performed, number of detected security errors, and execution times in second}
	\label{table:result1}
\begin{tabular}{|p{.5cm}|p{1.1cm}|p{1.1cm}|p{1.1cm}|p{1.1cm}|p{1cm}|}
	\hline 
App.&\# XSS tests&\# SQL tests&\# XSS detection&\# SQL detection&time(s)\\ 
	\hline 
1&1610&199&1&0&14\\ 
2&12358&796&0&0&924  \\ 
3&8209&398&10&4&29  \\ 
4&7347&199&0&0&81  \\ 
5&2527&398&3&0&1137  \\ 
6&5884&597&1&1&30  \\ 
7&9954&1194&1&0&49  \\ 
8&2464&796&0&0&1478  \\ 
9&1709&796&0&0&47  \\ 
10&16441&796&0&0&93  \\ 
	\hline 
\end{tabular} 
\end{table}

\subsection{Q2: Can the generated LTL properties detect incorrect implementation of patterns?}
\subsubsection*{Procedure}
To investigate Q2, the PHP applications were instrumented with the  debugger Xdebug, and we  collected logs composed of method call traces while the test case execution. Then, we used the tool Texada to check whether every LTL property holds in these method call traces.  When the pattern is strictly implemented as it is described in the UML sequence diagram of Figure \ref{fig:case} (1 class Validator), 4 LTL properties are generated, as in Table \ref{tab:rules}. However, the number of LTL properties may differ from one application to the other, with regard to the number of classes used to implement the security pattern. When there are more than 4 LTL properties for an application, the additional ones  capture the call of supplementary Validator classes and only differ from the properties of Table \ref{tab:rules} by the modification of the variable ValidatorURL. To keep our results comparable from one application to another, we denote with the notation $p_i$ the set of properties related to the property $p_i$ in \ref{tab:rules}.

Furthermore, both authors independently checked the validation part in every applications to assess how the security pattern is implemented in order to ensure that a property violation implies that a security pattern behaviour is not correctly implemented.

%difficuklty adapt sequence diagram ?

\subsubsection*{Results}
Table \ref{table:result2} lists in columns 2-5 the violations of the properties derived from those given in Table \ref{tab:rules} for the 10 applications. These results firstly show that our approach detects that the security pattern Intercepting Validator is never correctly implemented. The pattern seems to be almost implemented in App. 2 because only $p_4$ does not hold here. An inspection of the application code confirms that the pattern structure is correctly implemented as well as most of its method call sequences. But we observed that the application does not always return an error to the user when some inputs are not validated. This contradicts one of the pattern purposes.

App. 3, 4, 7-10 include some sorts of input filtering processes at least defined in one class. But, these do not respect the security pattern behaviours. Most of the time, we observed that the validation process is implemented in a single class instead of having an Intercepting Validator calling other Validator classes. This misbehaviour is detected by the violations of the properties  $p_2$ and $p_3$. Besides, we observed that the input validation is not systematically performed in App. 1, 5 and 6. This is detected by our tool with the violation of $p_1$. As a consequence, it is not surprising to observe that these applications are vulnerable to malicious injections. We also observed that when App. 5 validates the inputs, it does not  define the validation logic in a class. The fact that the security pattern is not invoked is detected by the violation of $p_1$. But, this property violation does not reveal that there is another validation process implemented. %% par les tests

In summary, our application code inspections confirmed the results of Table \ref{table:result2}. In addition to assessing whether the security pattern behaviours are correctly implemented, we observed that our approach may also help learn more information about the validation process, without inspecting the code. For instance, the properties based on $p_1$ check whether a validation method defined in a class is called every time a client request is received. The properties based on $p_4$ give information about the error management. Their violations express that users are not always warned when invalid inputs are provided to the applications.

\begin{table}[h]
	\caption{Results of the security pattern testing stage: violation of the LTL properties  and execution times in second}
	\label{table:result2}
	\centering 
	\begin{tabular}{|c|c|c|c|c|c|}
	\hline 
App.&$p_1$  &$p_2$  &$p_3$  &$p_4$  &Time(min) \\ 
	\hline 
1  &X&X&  &  &4,02 \\ 
2  &  &  &  &X&51,15  \\ 
3  &  &  & X& X&19,12  \\ 
4  &  &  &X&X&29,34 \\ 
5&X&X&X&X&6,5  \\ 
6&X&X&X&X&14,40  \\ 
7&  &  &X&X&24,77  \\ 
8&  &X&X&X&7,24  \\ 
9&  &X&X&X&5,56  \\ 
10&  &  &X&X&67,03  \\ 
\hline 
\end{tabular} 
\end{table}

\subsection{Q3: How long does it take to discover errors (Performance)?}
	
%How does our approach scale with the size of the test cases and of the traces?

\subsubsection*{Procedure}
We measured the time consumed by the tool to carry out security testing and  security pattern verification for the 10 applications. Execution times are given in Tables \ref{table:result1} and \ref{table:result2}. Furthermore, we also measured the number of LTL properties that are generated for 11 security patterns, which are often used with Web applications, as the LTL property number influences execution times.

\subsubsection*{Results}

\begin{figure}
	\centering
	\includegraphics[width=1\linewidth]{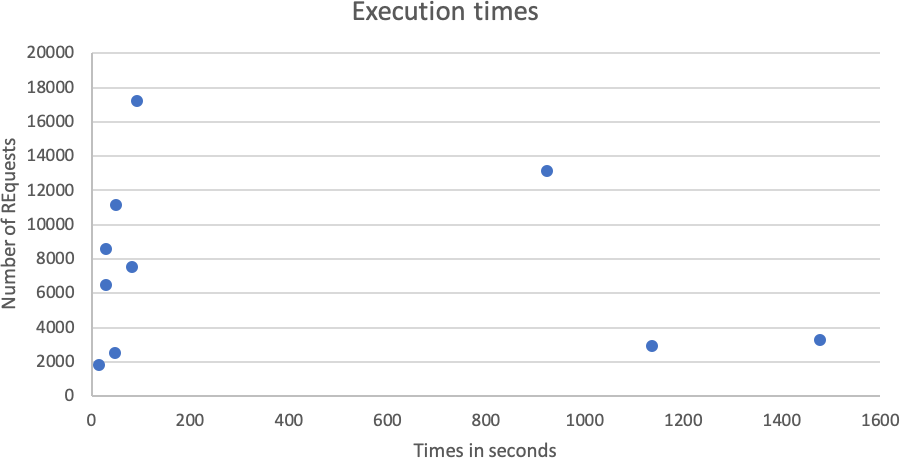}
	\caption{Execution times of the security testing stage for the ten applications}
	\label{fig:exec}
	%\vspace{-.5cm}
\end{figure}

The plot chart of Figure \ref{fig:exec} shows that security testing requires less than 2 minutes for 7 applications independently on the number of tests, whereas it requires more than 15 minutes for the 3 others. The security testing stage depends on several external factors, which makes it difficult to draw consistent conclusions. It firstly depends on the test case implementation; in our evaluation, we choose to call a penetration testing tool, therefore, execution times mostly depend on it. Another factor is the application structure (nb of classes, calls of external URLS, etc.). Therefore, we can only conclude here is that execution times are lower than 25 minutes, which remains reasonable with regard to the number of requests sent to applications.

The time required to detect property violations in method call traces is given in Column 6 of Table \ref{table:result2}. Execution times vary here between 4 and 67 minutes according to the number of traces collected from the application and the number of generated LTL properties. For example, for App. 1O, 17237 security tests have been executed, and 17237 traces of about 30 events have been stored in several log files. Furthermore, 7 LTL properties have been generated for this applications. These results, and particularly the size of the trace set, explain the time required to check whether the LTL properties hold. In general terms, we consider that execution times remain reasonable with regard to the trace set sizes of the applications. 

Table \ref{table:result3} finally shows the number of LTL properties generated from generic UML properties (without adapting them to application contexts) for 11 security patterns whose descriptions include UML sequence diagrams. For these patterns, the property number is lower or equal than 13. For every pattern, the property number is in a range that seems reasonably well supported by model checkers. However, if several security patterns have to be tested, the property number might quickly exceed the model-checker limits. This is we have chosen in our approach to check the satisfiability of each LTL property, one after the other, on method call traces.

\begin{table}[h]
	\caption{LTL property generation for some security patterns}
	\label{table:result3}
	\centering 
\begin{tabular}{|c|c|c|}
	\hline 
	Security pattern&\# UML diag. &\# LTL properties  \\ 
	\hline 
	Authentication Enforcer & 3 & 9  \\ 
	\hline 
	Authorization Enforcer & 3 &  13  \\ 
	\hline 
	Intercepting Validator & 2  & 4  \\ 
	\hline 
	Secure Base Action & 2 & 5  \\ 
	\hline 
	Secure Logger & 2 &  5  \\ 
	\hline 
	Secure Pipe & 2 &  10  \\ 
	\hline 
	Secure Service Proxy & 2 &  6  \\ 
	\hline 
	Intercepting Web Agent & 2 &  9   \\ 
	\hline 
	Audit Interceptor & 2 &  7   \\ 
	\hline 
	Container Managed Security & 2 &  7   \\ % Alternative/Implémentation d'un output controller %
	\hline 
	Obfuscated Transfer Object & 2 &  10   \\ 
	\hline 
	Obfuscated Transfer Object & 2 &  10  \\ 
	\hline 
\end{tabular} 
\end{table}

\subsection{Threat to Validity}
%We did not verify whether our method is applicable to any type of system. Therefore, the case study results cannot be generalized. 

This preliminary experimental evaluation is applied on 10 Web applications, and not on other kinds of software or systems. This is a threat to external validity, and this is why we avoid drawing any general conclusion. But, we believe that this threat is somewhat mitigated by our choice of application, as the Web application context is a rich field in great demand in the software industry. Web applications also expose a lot of well-known vulnerabilities, which helps in the  experiment set-up. In addition, the numbers of security patterns considered in the evaluation were insufficient. Hence, it is possible that our method is not applicable to all security patterns. In particular, we assume that generic UML sequence diagrams are provided in the security pattern descriptions. This is the case for the patterns available in the catalogue of Yskout et al. \cite{Yskoutcatalog}, but not for all the patterns listed in \cite{patrepo}. To generalise the approach, we also need to consider more general patterns and employ large-scale examples.

The evaluation is based on the work of students, but this public is sometimes considered as a bias in evaluations. Students are usually not yet meticulous on the security solution implementation, and as we wished to experiment vulnerable applications to check that our approach can detect security flaws, we consider that applications developed by students meet our needs. 

A threat to internal validity is related to the test case development. Our approach aims at guiding developers in the test case writing and security pattern choice. In the evaluation, we chose to complete test cases with the call of a penetration testing tool. The testing results would be different  (better or worse) with other test cases. Significant advances have been made in these tools, which are more and more employed in the industry. Therefore, we believe that their use  and the test cases considered in the experiments are close to real use cases. In the same way, we manually updated UML sequence diagrams to generate LTL properties that correspond to the application contexts. But, it is possible that we inadvertently made some mistakes, which led to false positives. To avoid this bias, we manually checked the correctness of the results by replaying the counterexamples returned by the model-checker and by inspecting the application codes.

\section{Conclusion}
\label{sec:conclusion}

Securing software requires that developers acquire a lot of expertise in various stages of software engineering, e.g., in security design, or in testing. To help them in these tasks, we have proposed an approach based on the notion of knowledge base, which helps developers in the implementation of secure applications through steps covering threat modelling, security pattern choice, security testing and the verification of security pattern behavioural properties. This paper proposes two main contributions. It assists developers in the writing of concrete security test cases and ADTrees. It also checks whether security patterns properties are met in application traces by automatically generating LTL properties from the UML sequence diagrams that express the behaviours of patterns. Therefore, the approach does not require developers to have skills in (formal) modelling or in formal methods. We have implemented this approach in a tool prototype \cite{data}. We conducted an evaluation of our approach on ten Web applications, which suggests that it can be used in practice.% on this kind of applications.

Future work should complement the evaluation to confirm that the approach can be applied on more kinds of applications. We also mentioned that security pattern descriptions do not all include UML sequence diagrams, which are yet mandatory by our approach. We will try to solve this lack of documentation by investigating whether security pattern behavioural properties could be expressed differently, e.g., with annotations added inside application codes. In addition, we intend to consider how our ADTree generation could support the teaching of security testing and security by design.

\bibliographystyle{IEEEtran}
\bibliography{doc}

\end{document}